\documentclass{article}
\usepackage[english]{babel}
\usepackage{authblk}

\usepackage[letterpaper,top=2cm,bottom=2cm,left=3cm,right=3cm,marginparwidth=1.75cm]{geometry}

\usepackage{amsmath}
\usepackage{graphicx}
\usepackage[colorlinks=true, allcolors=blue]{hyperref}

\title{\bf{City riots fed by \\ transnational and trans-topic web-of-influence}}
\author{Akshay Verma, Richard Sear, Nicholas J. Restrepo, Neil F. Johnson}
\affil{Dynamic Online Networks Laboratory, George Washington University, \\ Washington D.C. 20052, U.S.A.}

\begin{document}
\maketitle

\begin{abstract}
The sudden emergence of large-scale riots in otherwise unconnected cities across the UK in summer 2024 came as a shock for both government officials and citizens \cite{bbc2024ukriots,ofcom2024letter}. Irrespective of these riots' specific trigger, a key question is how the capacity for such widespread city rioting might be foreseen through some precursor behavior that flags an emerging appetite for such rioting at scale  \cite{isdglobal2024rumours,apnews2024britain,npr2024uk,time2024misinformation}. Here we show evidence that points toward particular online behavior which developed at scale well ahead of the riots, across the multi-platform landscape of hate/extremist communities. Our analysis of detailed multi-platform data reveals a web-of-influence that existed well before the riots, involving online hate and extremism communities locally, nationally, and globally. This web-of-influence fed would-be rioters in each city mainly through video platforms. This web-of-influence has a persistent resilience -- and hence still represents a significant local, national, and international threat in the future -- because of its feedback across regional-national-international scales and across topics such as immigration; and its use of multiple lesser-known platforms that put it beyond any single government or platform's reach.  
Going forward, our findings mean that if city administrators coordinate with each other across local-national-international divides, they can map this threat as we have done here and initiate deliberation programs that might then soften such pre-existing extremes at scale, perhaps using automated AI-based technology.

\end{abstract}

\section{Introduction}

The Office of Communication (Ofcom) which is the regulatory authority responsible for broadcasting and telecommunication in the UK, recently concluded that there was a clear and direct connection between the 2024 UK riots and activity on social media and messaging apps \cite{bbc2024ukriots}. However, their evidence was based on small sets of specific cases, leaving open the question as to how this arose at scale. In particular, how did the riots manage to mobilize thousands of people in otherwise unconnected cities across the UK virtually simultaneously \cite{isdglobal2024rumours,cnn_farright_2024,guardian_riots_2024}? And why did these riots escalate so quickly  into violence against mosques and migrant accommodation, coupled with the involvement of extremist groups such as Patriotic Alternative and the English Defense League \cite{nytimes_riots_2024}?

An initial, seemingly plausible explanation was that rioters, driven by the same misinformation about a Muslim asylum seeker, independently decided to take action in their respective cities \cite{apnews2024britain} \cite{npr2024uk} \cite{time2024misinformation}. However, this fails to explain the effectively simultaneous onsets of the disparate riots  and the similar speed of their escalations \cite{nytimes2024riots}. 
Instead, the in-depth Ofcom investigation suggests that the riots' near-simultaneous appearance across otherwise unconnected cities was driven by strategic use of social media by far-right groups. It is known that groups contributing to the riots, including remnants of the English Defence League (EDL), Patriotic Alternative, and Britain First, utilized platforms such as Telegram to coordinate actions and disseminate inflammatory content. According to {\em Le Monde}'s editorial on the morning of 17 August, 2024, `{\em Emeutes au Royaume-Uni : $\ll$L’extrême droite britannique est une constellation de groupuscules, sans véritable organisation ni hiérarchie$\gg$}' which means that the decentralized nature of these groups is like a “constellation of groupuscules without real organization or hierarchy,” which helped them collectively show swift mobilization and adaptability in their tactics. But to date, their online ecosystem has not been mapped out rigorously in a scientific way to test such a hypothesis.

The future threat of widespread city rioting is a big problem more generally. Other countries in Europe likely have something similar to the UK brewing under their societal surface, particularly given similar issues such as immigration in the framework of ongoing elections -- hence their cities are also likely vulnerable to such widespread riots. But again, without any evidence-based understanding of the underlying online ecosystem, how can any proper planning or perhaps precursors of such offline riots be identified in a scientific way? These questions provide the motivation for our study. We do not attempt here to quantify the specific amount of causality in terms of online behavior ahead of offline riots, but point instead to the weight of case-study evidence in the Ofcom enquiry which concluded that social media played a significant role in these widespread city riots, hence the term `fed' in the title of this paper.

In this paper, we present the most in-depth analysis to date of this online ecosystem prior to the UK's summer 2024 riots. Though we focus on the period before these specific riots, we see no reason why our findings would not be more generally applicable to other European cities and beyond, given the similar ways in which people use social media in cities globally and given the continued existence of the transnational and trans-topic web-of-influence that our study reveals. Our study is novel because of its depth in terms of the number of social media platforms and sheer number of communities that we include in our study. 

We stress that our research is not aimed at providing a quantitative study of the prediction of riots in general. However, by choosing to look in unprecedented depth and scale across an unprecedented number of platforms in the period before the UK's summer 2024 riots, our study does serve to complement existing data/algorithmic machinery mentioned as possibly predicting riots, and the studies that have used such tools \cite{Qiao2017} \cite{Galla2018} \cite{Korkmaz2016} \cite{10.1145/2623330.2623373} \cite{Warnke2024}. Examples of these existing tools are as follows: 

\begin{enumerate}
    \item {\em GDELT} (Global Database of Events, Language, and Tone, see \url{www.gdeltproject.org/}) monitors news sources worldwide, extracting details about events, actors, and sentiment. However, it has limited social media data.
    \item {\em EventRegistry} (\url{eventregistry.org/}) uses AI to track, analyze, and structure global news articles. It focuses on structured news, so social media tracking is limited.
    \item {\em ACLED} (Armed Conflict Location and Event Data Project, see \url{acleddata.com/}) focuses on conflict events, protests, and political violence. It appears to have little social media integration.
    \item {\em ICEWS} (Integrated Crisis Early Warning System) is a DARPA-funded project that collects and analyzes global event data for crisis prediction. However, it is not publicly accessible and it focuses mainly on political and security crises.
    \item {\em CrisisNET} was designed for humanitarian response and crisis management. It collects and organizes crisis-related data from global sources, including social media and news. It is no longer actively maintained, and it is not reliable for broad event tracking beyond humanitarian crises.
    \item {\em Talkwalker} (\url{www.talkwalker.com/}) tracks social media, news, and web sources for real-time event analysis. While good for brand reputation tracking, it is not ideal for security/conflict monitoring; it lacks a comprehensive historical archive.
    \item {\em Media Cloud} (\url{mediacloud.org/}) is an open-source platform that analyzes global media narratives, tracking keyword trends and topic framing. It comprises mostly mainstream news and has limited social media coverage.
    \item {\em Recorded Future} (\url{www.recordedfuture.com/}) uses AI to monitor cyber threats, geopolitical risks, and real-world events. However, this paid tool is not designed for protest or crisis tracking.
    \item {\em Sentinel Hub} (\url{www.sentinel-hub.com/}) provides access to satellite imagery and geospatial analysis for real-world event tracking.
    \item {\em Graphika} (see \url{graphika.com}) is a company that performs social media network analysis of topics such as extremism, radicalization, and influence operations. It does this by analyzing clusters of activity across social media. It focuses only on well-known platforms Twitter, Facebook, YouTube, Telegram, (as well as some niche forums) but does not encompass the large number included in the present study. 
    \item {\em Pulsar} (see \url{www.pulsarplatform.com}) focuses on social listening and audience analysis including tracking trending topics, audience behavior, and sentiment shifts across multiple platforms. Again, it does not have the large universe of platforms in our study.
    \item {\em EMBERS} (Early Model-Based Event Recognition using Surrogates) is a tool that focuses on civil unrest forecasting using multiple open-source indicators including social media, economic indicators, and news reports, to detect emerging threats and predict unrest before it happens. It is not publicly accessible -- however, we know from the involvement of one of us in the research program where this tool originated, that it does not include the range of platforms included in our present study.
Moreover, {\em EMBERS}'s actual performance success is not publicly known. 
   \item There are also statistical studies which try to predict riots based purely on the analysis of time series. We note that the success rate of such pure time-series analyses is open to question; and even if/when they get some right, it will be hard to know why since the predictions are not based on a core understanding of what mobilized the rioters.
\end{enumerate}

To summarize our paper's main contribution, this study reveals the detailed, city-centric online infrastructure that existed before the UK riots. This infrastructure would have enabled the rapid spread of extremist narratives and real-world mobilization at scale. Specifically, we uncover a trans-national and trans-topic infrastructure that was in place prior to the riots, and which fed the would-be rioters. We find this by mapping at unprecedented resolution the online built-in communities across platforms that are associated with the Patriotic Alternative (PA): a British far-right, neo-Nazi, and white nationalist group \cite{gnet2021biologised} founded by Mark Collett in 2019. We then look who they link to and who links to them; i.e. any member of such a PA community (community 1) can at any time cross-post content of interest from another community (community 2) and hence create a link from 1 to 2 (see Fig. 1A). Other members of 1 are then alerted to 2’s existence, and can visit community 2 to share their content and thoughts. Reference \cite{velasquez_online_2021} provides explicit
examples of such links and communities (nodes). Our paper is organized as follows: Section 2 describes the data that we used, which is available online and whose collection methodology has been used and described extensively in prior publications, e.g. Refs.  \cite{Zheng_Sear_Illari_Restrepo_Johnson_2024,velasquez_online_2021}. Section 3 describes our findings, Sec. 4 provides discussion, and Sec. 5 explains our conclusions.

\vskip0.1in
Our findings mean that if city administrators coordinate with each other across local-national-international divides, they can map this threat as we have done here and initiate deliberation programs that might then soften such pre-existing extremes at scale, using automated AI-centric technology as in Refs. \cite{softeningextremes,scienceatrevete}. We discuss this further in Sec. 5.

\section{Data}

Our dataset used in this study is available at \url{https://github.com/gwdonlab/data-access}. Due to the sensitive nature of the original social media data, direct sharing of raw datasets is restricted to comply with data protection standards and prevent misuse. However, processed derivative datasets are available for reproducing the study’s results.
Our dataset is collected continuously in near-real-time across approximately 30 platforms using a hybrid methodology that combines machine-learning and expert human oversight, as described in detail in various prior works \cite{Zheng_Sear_Illari_Restrepo_Johnson_2024,velasquez_online_2021,lupu_offline_2023}. We will briefly review this data collection methodology here, referring to Refs. \cite{Zheng_Sear_Illari_Restrepo_Johnson_2024,velasquez_online_2021,lupu_offline_2023} for full details.

Our main units of analysis are the social media communities which can be created by users on each platform, e.g. a Club on VKontakte; a Page on Facebook; a Channel on Telegram; a Group on Gab. It is known that people join these communities to develop their shared interests, to express their feelings with similarly minded individuals (including their anger and hence potentially hate and extremism) and that this can facilitate mobilization around issues \cite{Zheng_Sear_Illari_Restrepo_Johnson_2024,velasquez_online_2021,lupu_offline_2023}. Each such in-built community (which for simplicity we henceforth refer to simply as a community) is a well-defined entity with its own unique ID assigned by the platform; it is completely unrelated to the term `community' used in network analysis to detect structural clusters; and it contains anywhere from a few to a few million users \cite{Zheng_Sear_Illari_Restrepo_Johnson_2024,velasquez_online_2021,lupu_offline_2023}. We define a `hate community' as one where our subject matter experts determine that at least two of the most recent twenty posts (at the time of analysis) exhibit hate and related extremism, which we classify based on the FBI’s definition of hate crimes and content promoting extreme racial identitarianism or fascist ideologies \cite{Zheng_Sear_Illari_Restrepo_Johnson_2024,velasquez_online_2021,lupu_offline_2023}. 

As indicated in Fig. 1A, a link is formed from a particular hate community (e.g. community 1, source node) to another community (e.g. community 2, target node) when community 1 shares a URL with its members that links directly to community 2 or to content posted in community 2. Such a link serves to direct the attention of community 1’s members to community 2, which may be on a different platform and may even be in another language. Community 1’s members may then add comments and content in community 2 without community 2’s members knowing this incoming link exists. Hence, community 2’s members -- of which there may be tens of thousands or more -- can unwittingly experience direct exposure to, and influence from, community 1’s hateful narratives. Community 1's members may also include a link to their own community, encouraging community 2's members to join community 1.

These links result in a directed network where hate communities are the source nodes. If a target node is not classified as a hate community, it is labeled as “vulnerable mainstream” because it is just one link away and hence at high risk of seeing or adopting  hateful rhetoric from the source node (which is always a hate community in our study; we do not
include links originating in vulnerable mainstream nodes because our focus is on hate networks). This is consistent with empirical research in social psychology which shows that contempt-based in-group superiority strongly predicts hate speech \cite{Bilewicz_Kaminska_Winiewski_Soral_2017}.

Mapping out all these communities and links across all platforms as in Fig. 1A reveals a complex web (each node is a community, each directed link is a URL from one node to another) that features hate/extremism content. We refer to this for brevity as the `hate universe'. Since links appear over time, and can get lost, removed, or forgotten, this network is dynamic (Fig. 1A) and these dynamics can facilitate exposure and engagement while evading moderators, as well as opening up users to new pathways to new hate/extremism content and communities. Explicit examples of these communities and links are documented in our prior papers (e.g. Refs. \cite{lupu_offline_2023,Zheng_Sear_Illari_Restrepo_Johnson_2024,velasquez_online_2021}).

Our results are obtained from applying this same methodology in order to identify the online presence of the Patriotic Alternative (PA); i.e. we focus on the subset of PA nodes in our hate universe (Fig. 1B). Each node's geographic identifiers (such as city and region names) were assigned coordinates, while nodes without specific regional identifiers were categorized at the national level. All our findings and subsequent discussion in  this paper are based on this particular PA subset of our hate universe. For convenience, we aggregate it over time for the period prior to the summer 2024 riots. One could use our dataset at a more granular level to look at the daily link formation, but we leave this for a future study since it does not change the pre-riot existence of the PA web that we discuss.

While no social media dataset can be entirely comprehensive, this hate universe encompasses billions of individual accounts and is the largest mapping that exists. We do not know for sure that each individual account is a specific individual, nor do we know how much overlap there is between specific people and their accounts across multiple platforms, but for simplicity we use the term ``individual" together with approximate numbers since the precise number of real individuals is not relevant to our findings concerning the web-of-influence itself.

\begin{figure}
\centering
\includegraphics[width=1.0\linewidth]{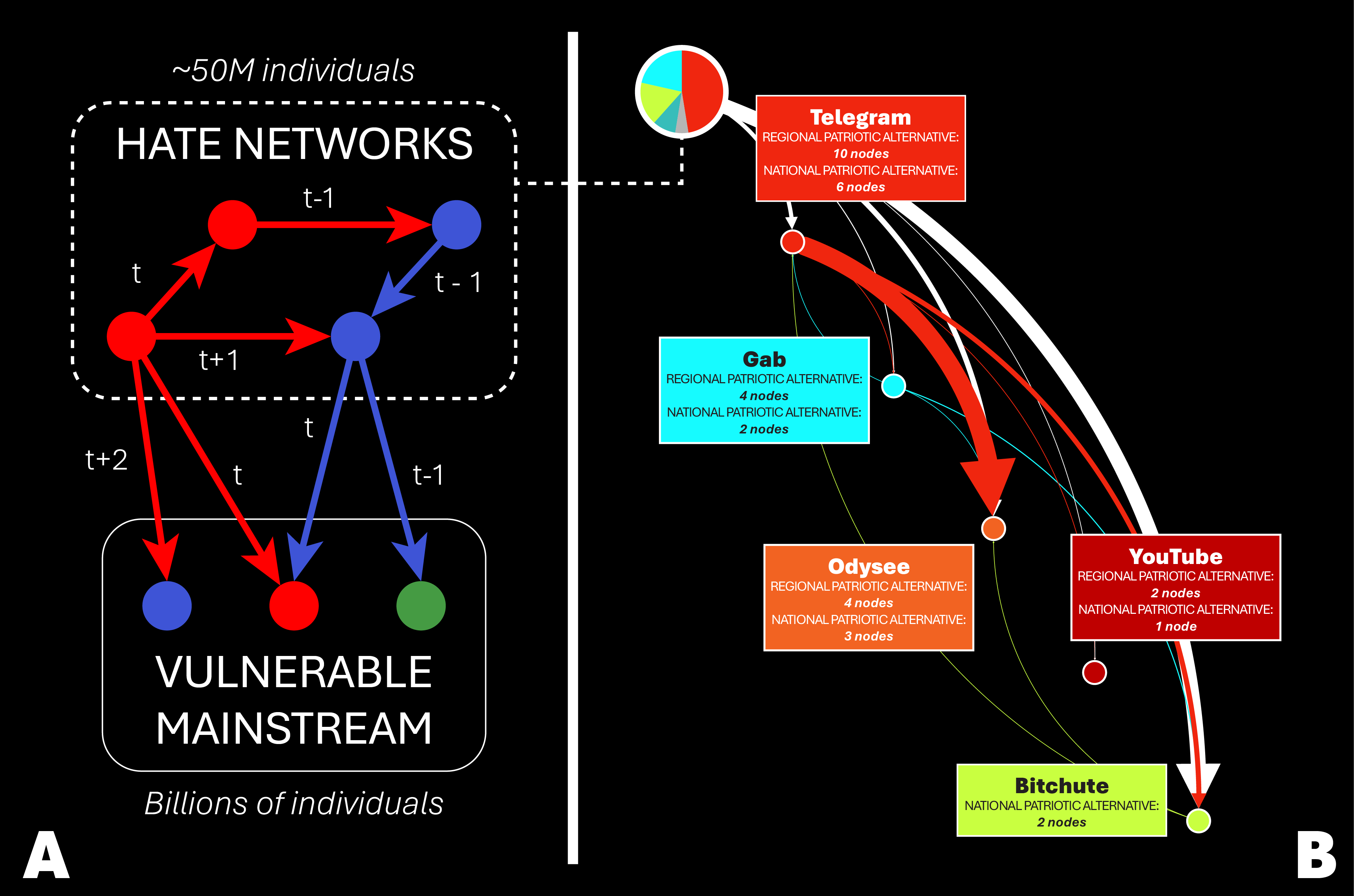}
\caption{
A: Conceptual diagram of how communities in the global hate network link to each other and to communities outside the network over time. B: The broader hate network's many connections to the Patriotic Alternative's local and national channels. Both A and B indicate snippets of the (huge) hate universe that we collected, as discussed in Sec. 2.
 }
\label{fig:PA_in_HU}
\end{figure}

\section{Findings}
Our hate universe dataset shows that during the two years before the summer 2024 riots, Patriotic Alternative (PA) established a decentralized yet highly interconnected network of regional city-based chapters (e.g., Patriotic Alternative London, Patriotic Alternative Yorkshire) within the broader online ecosystem (Figs. 1B,2). Each city-based chapter is a node in the hate universe. Focusing on this PA node subset, we see that these city-level PA nodes are not isolated. Rather, they are linked to national PA hubs (Fig. 2 upper right side) where localized grievances are consolidated into overarching extremist narratives. This process helps foster a sense of collective struggle across different cities, allowing discontent in one location to quickly resonate across others.

Digging into its structure across platforms reveals that PA strategically leveraged platforms with relatively little moderation (e.g. Telegram, Odysee, and Gab) in  order to expand its reach, using these particular networks to coordinate actions while evading platform moderation. Our findings are also consistent with Ofcom’s conclusion that closed groups on messaging apps circulated content identifying targets for violence during the riots \cite{ofcom2024letter}. However, our study significantly extends Ofcom's verbal narrative by demonstrating how PA’s structure facilitated the rapid coordination of unrest: specifically, city-based PA nodes amplified local grievances, which were then reinforced by national PA nodes hence creating a feedback loop of radicalization and mobilization (Fig. 2 upper right side).

More worryingly for cities worldwide, PA is also embedded in a transnational far-right network, linking British extremists to international white nationalist and conspiracy-driven communities (Fig. 2 left side and bottom). These global far-right nodes serve as echo chambers, validating and amplifying the grievances expressed by city-based PA groups \cite{goel2023hatemongers}. This international reinforcement provides ideological justification for local extremism, making city-based unrest not just a reaction to domestic events but part of a broader, global far-right movement \cite{liang2022far}. By understanding the structure of PA and its relation to these larger networks, we gain critical insights into how online extremism continues to transcend localized politics and hate.

\subsection{Patriotic Alternative (PA)'s Web}

The structure of the Patriotic Alternative (PA) web embedded within the online hate universe, reveals that the network is highly centralized around three major PA entities: `Laura Towler', `Mark Collett', and the Patriotic Alternative Organization. Mark Collett is the founder of Patriotic Alternative, and Laura Towler is the deputy leader and one of the organization's most frequent spokespersons. The online communities centered around these entities form the national-level PA nodes, as they are not tied to specific regions or cities but instead function as central conduits for extremist narratives and hate speech.

Each of the three entities maintains a nationwide presence across multiple unmoderated platforms: Telegram, BitChute, Gab, and Odysee. `Mark Collet' has primary channels on Telegram and Odysee all named after him. `Laura Towler' uses the same platforms and has primary channels, similarly named after her, on each of them. Patriotic Alternative, the organization, also uses the same platforms and has its own channels across them. It is of course possible that other people operate their communities and also set these communities up, and we are not in any way claiming that these specific individuals are themselves  responsible. But the end result remains the same: the perception is that these communities and their activity are largely associated with them and hence effectively act as them.

\begin{figure}

\centering
\includegraphics[width=1.05\linewidth]{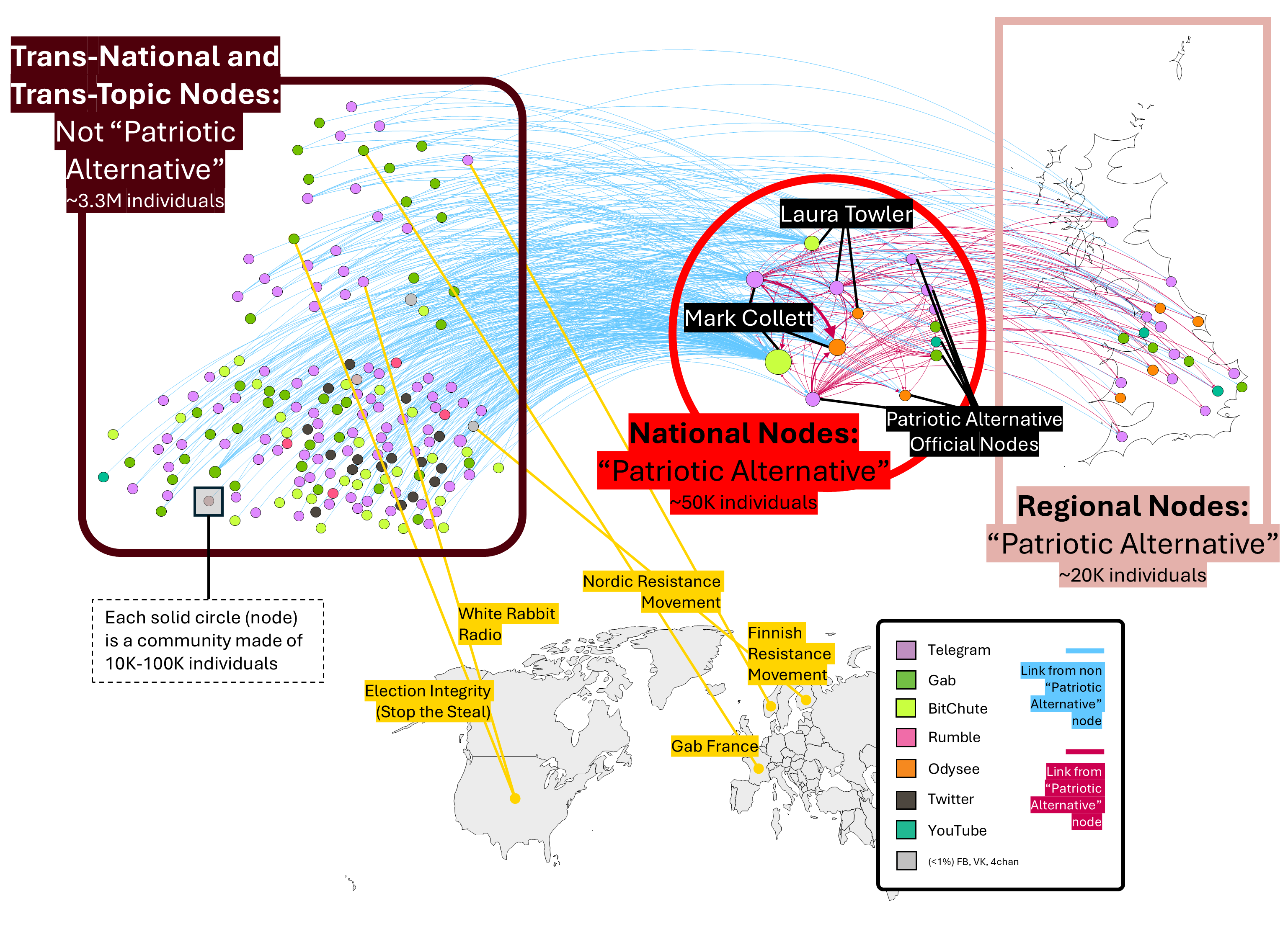}
\caption{ 
The `City Amplification Network' (data aggregated for simplicity from 2020 to just before the riots in summer 2024). This visualization shows how Patriotic Alternative (PA) operates online communities across various cities and regions in the UK, which feed information to the major PA nodes and which in turn are connected to trans-national and trans-topical nodes.
 }
\label{fig:city_amplification_network}
\end{figure}

Beyond its national-level structure, PA operates a city-centric network, which we characterize as the `City Amplification Network', seen in Fig. \ref{fig:city_amplification_network}. This structure consists of multiple regional and city-based PA chapters (nodes) such as Patriotic Alternative Scotland, Patriotic Alternative London, and Patriotic Alternative East Midlands. These localized communities (nodes) function as regional hubs, gathering and curating extremist content before funneling it upward to the national PA network (Fig. 2 upper middle).

This bottom-up information flow is crucial to PA’s ability to maintain a cohesive yet decentralized presence across the UK. Local city-based nodes generate and circulate content, which is then amplified and validated at the national level by `Collett', `Towler' and the broader PA organization. This mechanism not only consolidates narratives across cities, but also fosters a shared sense of grievance and unity, making it easier for PA to coordinate large-scale mobilization, as observed in the 2024 riots.

PA strategically leverages multiple social media platforms, with a significant emphasis on unmoderated spaces that allow extremist content to spread without restriction. Our analysis finds that PA has a notable presence on Telegram (16 accounts), Odysee (7 accounts), and Gab (6 accounts). Platforms like Odysee and BitChute serve as key hubs for video-based propaganda.

When we examine the digital infrastructure of regional PA chapters, we see a similar pattern. 10 out of 20 city-based PA chapters (nodes) rely on Telegram, 4 using Odysee and 4 using Gab. Furthermore, when analyzing hate communities (nodes) targeting these same cities, 18 out of 24 show a strong presence on Telegram, with an additional 5 on Gab. This underscores the critical role of unmoderated platforms, particularly Telegram and Gab, in connecting UK cities to the broader national PA ecosystem and to international far-right movements.

\subsection{Relation to transnational nodes and trans-topic nodes}

In the broader hate network surrounding PA, we observe the emergence of several transnational and trans-topic nodes. These nodes extend beyond PA's immediate regional and national politics, connecting it to a wider network of far-right movements and conspiracy-driven communities across the globe. The transnational nodes represent far-right groups from other countries, while trans-topic nodes engage in discussions that shift across multiple extremist ideologies, such as anti-immigration, antisemitism, and conspiracy theories like election fraud [`Election Fraud Exposed'] and Anti-vaccine activism [`BewareTheNeedle – Anti-Vaxxers, Ex-Vaxxers, and Vaccine Risk-Curious']. The presence of these nodes highlights that PA is not operating in isolation; instead, it is embedded within a larger global ecosystem of hate, potential mis/disinformation and influence. These international and multi-thematic connections allow PA to amplify its message, aligning with and benefitting from global waves of mis/disinformation, hate speech, and radicalization.

The existence of these transnational and trans-topic nodes suggests that violent protests and/or acts of hate in a given city may not be not confined to a single issue or region -- and may not even be relevant to that city. Instead violent protests and/or acts of hate in a given city can transcend both geography and subject matter, representing issues only in some other seemingly unrelated and possibly quite distant city. Nodes like `US Voter Fraud \& Coup-Ops Intel' and `The Reality Report', which were highly influential during the 2020 US presidential elections and the following Jan. 6 Capitol riots \cite{verma2024us}, and are also present in the PA pipeline, demonstrate that communities involved in one hate-driven topic or event in one city often readily engage in other seemingly unrelated topics which may be of little relevance to their city. This fluidity highlights how hate speech is not topic-specific, but instead part of a broader ecosystem where individuals and groups shift focus between hate narratives, depending on what is polarizing at the moment. Furthermore, the presence of Europe-based nodes like the `Nordic Resistance Movement' and US-based nodes like `White Rabbit Radio Live' \cite{splc2013white} illustrates how nationalist movements across borders fuel a global exchange of hate, mis/disinformation and extremism. These nodes reveal that hate, mis/disinformation and extremism thrive on a transnational scale and cross thematic lines from antisemitism, election denial, and anti-vaccination conspiracies to anti-immigration sentiments, blurring the boundaries of what motivates those eventually involved in on-street acts in a given city. 

The city-based PA chapters act as entry points into this wider transnational ecosystem. When Patriotic Alternative London, Patriotic Alternative Yorkshire, or Patriotic Alternative East Midlands communities (nodes) engage with these transnational and trans-topic communities, they funnel global extremist content into UK-specific grievances. This reinforces and intensifies radicalization at the local level, making national-level mobilization -- such as the 2024 UK riots -- more likely and more coordinated across multiple cities. Fig. 1B showcases how PA communities are just a tiny part of the larger `hate universe'.

\begin{figure}
\centering
\includegraphics[width=1.0\linewidth]{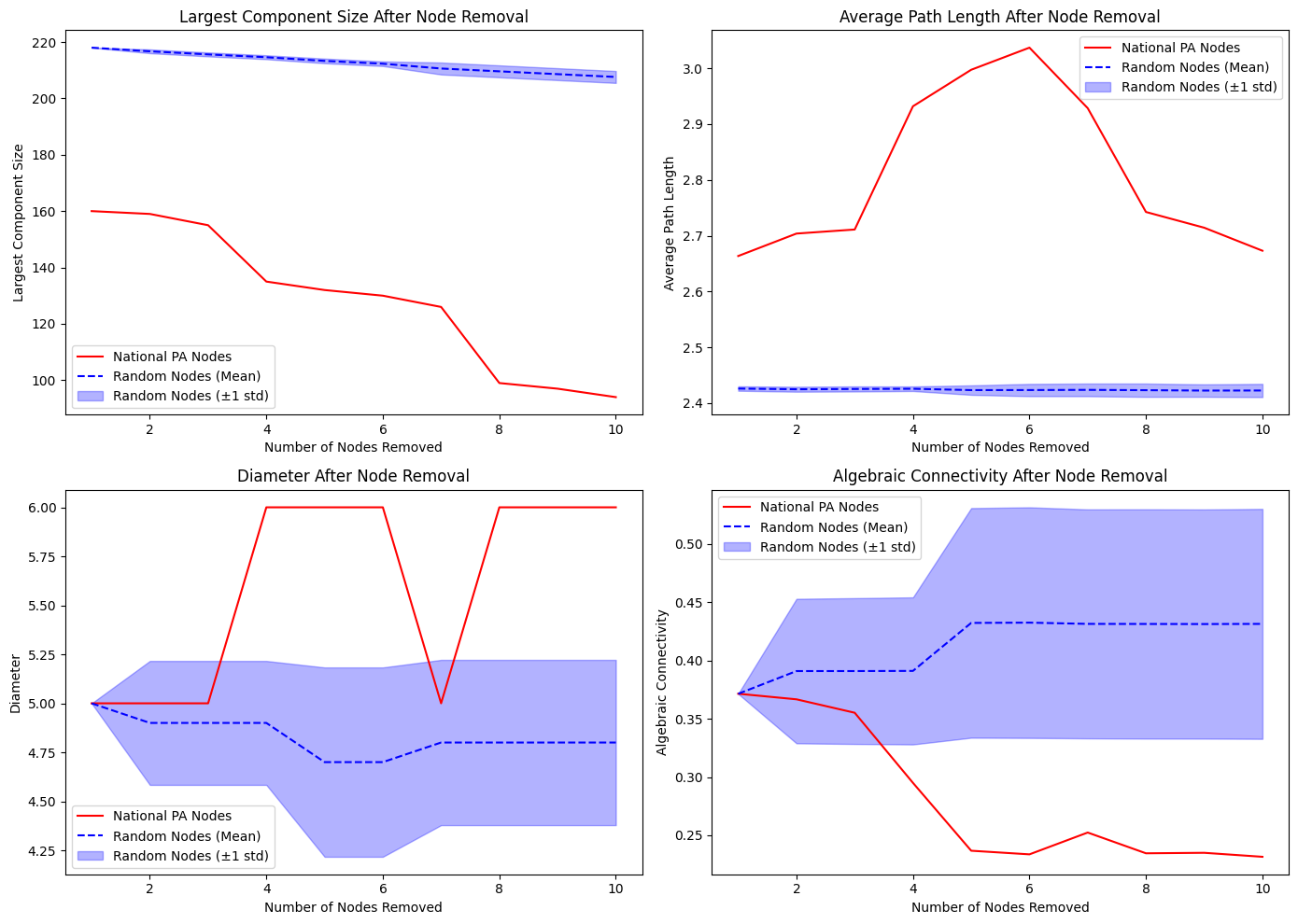}
\caption{Impact of Node Removal on Patriotic Alternative Infrastructure. These plots compare the effects of removing the National Patriotic Alternative nodes to removing nodes at random, through the impact of this removal on 4 key metrics: Largest Component Size, Average Path Length, Diameter, and Algebraic Connectivity. When National Patriotic Alternative nodes are removed, there is a significant impact on the connectivity of the hate/extremism network. In contrast, removing nodes at random does not significantly affect the overall connectivity of the network, indicating that its infrastructure is resilient to attacks of these types.
}
\label{fig:Network_Fragmentation}
\end{figure}

\section{Discussion}

Our findings extend beyond just describing the Patriotic Alternative network. For example, `Mr. AG' a Neo-Nazi who was active during the 2024 UK riots and who advised rioters on arson tactics \cite{bbc2024news}, was mentioned multiple times in a Nordic Resistance Movement chat which is part of the  PA pipeline revealed in this paper.  Additionally, the `PA Gaming' chat, of which `Mr. AG' was a part \cite{bbc2024news}, is also present in the pipeline, connected to the official Patriotic Alternative Telegram channel and to the `Mark Collett' Telegram channel. These instances corroborate how the infrastructure we describe in this paper can be used for mitigation of future hate events targeting cities.

Without an understanding of this infrastructure, it would be practically impossible for the relevant authorities -- whether government entities or platform operators -- to effectively disconnect UK cities from the national-international hate communities. Figure \ref{fig:Network_Fragmentation} illustrates this by comparing two node removal strategies: one where the infrastructure is known, allowing for the targeted removal of national PA nodes that serve as a medium between the international hate communities and UK cities, and another where nodes are removed at random without prior knowledge of the infrastructure in place. In both strategies, 10 nodes are removed sequentially, with the impact of removal measured for 4 network metrics: largest component size, average path length, network diameter, and algebraic connectivity. These measures give us insights into the connectivity and resilience of the network's infrastructure. Largest Component Size represents the size of the largest connected group of nodes within the network, so a decrease in this metric indicates that the network is fragmenting, with fewer nodes remaining interconnected. Average Path Length measures the average shortest distance between any two nodes; an increase in this length suggests that nodes are further apart on average, making communication across the network slower and more difficult. Diameter of the network, which is the longest shortest path between any two nodes, reflects the overall spread of the network: a larger diameter means the network is stretched out, requiring more steps to reach distant nodes. Algebraic Connectivity reflects the overall robustness of the network \cite{4231826}. A decrease in Algebraic Connectivity indicates that the network is becoming more vulnerable to fragmentation, as less disruption is required to break it into isolated parts.

These findings reveal that removing nodes at random does not significantly affect network connectivity or resilience across any of these four metrics. However, when only national PA nodes are removed (a targeted approach that requires prior knowledge of the infrastructure) there is a significant impact: the Average Path Length and Diameter increase, while the Largest Component Size and Algebraic Connectivity decrease. This shows that the network becomes less connected, with longer distances between nodes, and less resilient, as evidenced by the drop in algebraic connectivity. As such, this indicates what effective mitigation will require.

\section{Conclusions}

Our research demonstrates how UK cities are part of a regional-national-international infrastructure of hate communities. This infrastructure connects UK cities to national movements and beyond to the much broader transnational and trans-topical international hate universe. It forms a web-of-influence, weaving regional grievances and narratives with broader national and global ideologies, such as anti-immigration, neo-Nazism, election fraud conspiracies, and COVID-19 denialism. Collectively, these interconnected communities create a powerful hate ecosystem that facilitates the rapid dissemination of extremist narratives, amplifying and intensifying hate events within UK cities.

Our findings indicate that understanding this infrastructure can enable more effective prediction and mitigation of future hate-fueled incidents across the UK and comparable cities outside the UK. This web-of-influence targeting UK cities relies on a continuous flow of narratives that move both upward, from regional hate chapters to their national and international counterparts, and downward, where global ideologies reinforce local grievances. Our findings also show that this infrastructure is especially resilient because of its many connections to national-international influence and its use of multiple communities across many unmoderated platforms like Telegram, Gab, BitChute, and Odysee.

Our findings also mean that isolated attempts to contain hate events at a city level will fail to counter the larger, interconnected nature of these hate networks. Instead, any effective response should be collaborative, involving city administrators, national security agencies, and platform operators. Without a clear understanding of the cross-platform, cross-border structure of hate networks, local containment efforts will not affect the connectivity of the infrastructure. Coordinated actions across regional-national-international divides are essential to disconnect cities nationally and globally from the web of extremist influence that we identify, and to mitigate the risk of future hate-driven events, like the 2024 summer UK riots.

Looking toward a solution, our findings mean that if city administrators coordinate with each other across local-national-international boundaries, they can map this threat as we have done here and initiate deliberation programs that might then soften such pre-existing extremes at scale, perhaps using automated AI-centric technology. Specifically, it was shown empirically in Ref. \cite{softeningextremes} that automatically assembling groups of users online with
diverse opinions, guided by a map of the online social media infrastructure, and facilitating their anonymous
interactions, can lead to a softening of extreme views. A subsequent study \cite{scienceatrevete} then showed how AI could be used to scale this scheme, meaning that it could be run fairly continuously online at scale and in an automated manner, without necessitating
the contentious removal of specific communities, imposing censorship, relying on preventative messaging,
or requiring consensus within the online groups. Computer simulations confirm \cite{softeningextremes} that even if external variables occasionally
decrease the softening effect, the continual and parallel
operation of forming diverse anonymous groups can lead
to accumulative softening effects over time and at scale. All that is required is knowledge of the nodes and links from our hate universe dataset. 

\section{Data Availability}

Our dataset used in this study is available at \url{https://github.com/gwdonlab/data-access}. The original data contain sensitive information from social media platforms; to comply with data protection standards and avoid potential misuse, the raw data cannot be shared publicly. However, the processed derivative datasets can be used to reproduce the results in the study.

\section{Code Availability}

The code is available at \url{https://github.com/gwdonlab/data-access}.
The network visualizations in Fig. \ref{fig:city_amplification_network} was created using Gephi. Figures in Fig. \ref{fig:Network_Fragmentation} were generated in python using Matplotlib.
Together, this provides readers with access to the minimum dataset that is necessary to interpret, verify, and extend the research in the article. Cosmetic changes to figures were carried out in Adobe Illustrator, a commercially available piece of software.

\section{Acknowledgements}

N.F.J. is supported by US Air Force Office of Scientific Research awards FA9550-20-1-0382 and FA9550-20-1-0383, and by The John Templeton Foundation.

\section{Competing Interests}

The authors declare no competing interests.

\bibliographystyle{ieeetr}
\bibliography{main}

\end{document}